\documentstyle[12pt,aasms4,psfig]{article}

\singlespace

\begin{document}

\title{THE\  DECELERATION\  OF\  GIANT\  HERBIG-HARO\  FLOWS} 

\author{ Elisabete M. de Gouveia Dal Pino$^{1}$ 
 }
\altaffiltext{1}{Instituto Astron\^omico e Geof\'{\i}sico, Universidade
de S\~ao Paulo, Av. Miguel St\'efano, 4200, S\~ao Paulo
04301-904, SP, Brasil; 
E-mail: dalpino@iagusp.usp.br}

\vskip 2.0 cm


\begin{abstract}
It has been recently discovered that spatially separated Herbig-Haro objects, 
once considered unrelated, are linked within a chain that may extend for parsecs 
in either 
direction of the embedded protostar forming a {\it giant Herbig-Haro jet}.
Presently, several dozen of these giant flows have been
detected  and the best documented example, the HH~34 system, 
shows a systematic velocity 
decrease with distance on
either side of the source.
In this paper, we have modeled giant jets by
performing  fully three-dimensional simulations of
 overdense, radiatively cooling jets modulated
with long-period (P $\sim$ several hundred years)
and large amplitude sinusoidal velocity variability at
injection ($\Delta v \sim$ mean jet flow velocity).  Allowing them to
travel over a distance well beyond the source, we have found that
multiple travelling pulses develop and their 
 velocity indeed falls off smoothly and 
systematically with distance. This deceleration is fastest if the jet is 
pressure-confined,
in which case the falloff in velocity is roughly consistent with
the observations. The deceleration occurs as momentum is 
transferred by gas
expelled sideways from the traveling pulses.  

The simulation of a pressure-confined, steady-state jet with similar initial 
conditions to those of the pulsed jet 
shows that the flow in this case experiences $acceleration$. This result 
is thus an additional indication that the primary source of 
deceleration in the 
giant flows
$cannot$ be  attributed  to braking of the jet head 
against the external medium.


\end{abstract}

\clearpage

\section{Introduction}

The optical jets associated with low-mass, young stars have long been of
interest. (For a review, see Reipurth \& Raga 1999). First discovered over 15 
years ago (Mundt \& Fried 1983), these narrow strands of emission trace stellar
winds that have been collimated in a bipolar fashion. The discrete 
Herbig-Haro knots along the jets are radiating shock fronts within this flow. 
Their bow shock morphology, high spatial velocity, and symmetric pattern with 
respect to the star (e.g., Zinnecker et al. 1998) all indicate that the shocks
generally arise through the steepening of velocity fluctuations in the 
underlying, highly supersonic wind. Theoretical studies have confirmed that 
traveling shocks created in this manner reproduce the essential properties of 
the observed knots (Raga et al. 1990; Stone \& Norman 1993; de Gouveia Dal 
Pino \& Benz 1994).

Researchers at first considered the most luminous Herbig-Haro objects to be
terminal bow shocks, i.e., to mark the location where the stellar wind impacts
its cloud environment. However, these regions are typically displaced about
0.1~pc from the star, while the measured jet speeds are about 300~km~s$^{-1}$.
The corresponding travel time of 300~yr is a small fraction of the $10^5$~yr
period over which low-mass young stars are thought to drive vigorous winds 
(Adams, Lada, \& Shu 1987). With the recent availability of wide-field CCD 
arrays, it has become clear that jets indeed extend to far greater distances. 
Spatially separated Herbig-Haro objects, once considered unrelated, are now 
seen to be links within a chain that may continue for parsecs in either 
direction from the embedded, central object.

The first clearly identified {\it giant Herbig-Haro flow} was the HH~34 system
in the L1641 dark cloud of Orion~A (Bally \& Devine 1994). A Herbig-Haro object
of that name had long been known (Herbig 1974), as had a well-delineated
optical jet pointing back to the $45~L_\odot$ embedded source HH~34~IRS 
(Reipurth  1985). Wide-field imaging showed that about a dozen Herbig-Haro
objects fall along a gentle S-curve centered on the star, and extending over
1.5~pc. This linkage is not merely a visual impression. In a followup study,
Devine et al. (1997) showed that all of the Herbig-Haro objects to the north
of HH~34~IRS are redshifted, while the southern ones are all blueshifted. The
actual radial velocities, moreover, decrease smoothly and systematically on
either side of the star.

Since this initial discovery, several dozen additional giant flows have been
found. All broadly resemble HH~34, but with conspicuous variations. Thus, the
chain of Herbig-Haro objects centered on the HH~111 jet stretches nearly 8~pc
from end to end, but exhibits very little bending (Reipurth et al. 1997). 
While the HH~34 flow has a low level of molecular emission near its ends, 
other systems, such as RNO~43, are encased within massive amounts of cloud gas
and broad CO outflows (Bence et al. 1996). The HH~333 flow in the crowded 
field of NGC~1333 displays the usual S-curve, but is centered on an optically 
visible star (Bally et al. 1996).

These observations raise a number of important issues. How do jets maintain
collimation over such extraordinary distances? What role do they play in
dispersing the ambient cloud gas? What is the explanation for the S-shaped 
curves? Finally, why are the shocks slowing down?

On this paper, we embark on our own investigation of giant jets by focusing,
at least initially, on the last question. The best documented example of
deceleration is still the HH~34 jet, although the phenomenon has been seen in a
number of others, such as HH~83, HH~335, RV~Cep, and HH~111 (Reipurth et al.
1997). Figure 7 of Devine et al. (1997) shows that the radial velocities within
both the north and south lobes of HH~34 decrease in a nearly linear fashion
with distance, falling by at least a factor of 4 on either side of the
driving star. Moreover, the measured rates of velocity decline are strikingly 
similar: 110~km~s$^{-1}$~pc$^{-1}$ for the southern (blueshifted) lobe and 
137~km~s$^{-1}$~pc$^{-1}$ for the northern (redshifted) one. Proper motions
of the most prominent optical knots also exhibit a falloff, but the 
observational errors are too large for detailed comparison of the slopes.

The pattern of radial velocities argues that the deceleration is {\it not}
created by braking against the external medium, which can hardly be expected to
possess this degree of symmetry. In any case, the HH~34 jet projects into a
region that is relatively depleted in molecular gas, as evidenced by the weak
CO emission (Edwards \& Snell 1984, Reipurth et al. 1986, Chernin \& Masson
1995). The deceleration, therefore, must somehow be intrinsic to the flow
itself. Both the proper motions and radial velocities in HH~34 are higher than
the shock speeds deduced from emission lines (Bally \& Devine 1997). This is
the same pattern exhibited in shorter, contiguous jets, and demonstrates once
more that the chain of bow shocks arises from velocity fluctuations in the 
underlying wind. Under what circumstances, then, do the shocks in a
variable jet decelerate?

To address this question, we have modeled the jet through direct, numerical
simulation. The brightest bow shocks within the giant flow must be created by
order-unity changes of the wind velocity. Hence, our calculations simulate
a jet modulated with large amplitude, and we allow material to travel an
appropriate distance beyond the input source. The combination of large
amplitude and computational length distinguishes our study from previous
efforts. We find that the velocity near the central axis indeed falls
systematically. This deceleration is fastest if the jet is pressure-confined,
 in which case the falloff in velocity is roughly consistent with
the observations. The momentum lost by the axial matter is transferred to gas
expelled sideways from the traveling pulses. 

In Section 2 below, we detail the computational method employed in these
simulations. Section 3 then presents numerical results. We stress from
the outset that our study is limited in scope, and is not intended to
simulate all aspects of the giant flows,  nor to provide the coverage of
parameter space that would be appropriate in a fuller treatment. Nevertheless,
we feel that our calculations illustrate the basic principle that underlies the
observed deceleration. Finally, Section 4 discusses the observational and
theoretical implications of our findings.

\section{Computational Method}
Our goal has been to study the behavior of a pulsed, radiatively cooled jet
as it propagates into a medium of uniform density. For this purpose, we
utilized a fully three-dimensional smoothed-particle hydrodynamics (SPH) code
originally developed to investigate the working surfaces of steady jets
(de Gouveia Dal Pino \& Benz 1993). The code was subsequently employed, with
appropriate modifications, to study low-amplitude, pulsing jets (de Gouveia
Dal Pino \& Benz 1994), propagation into stratified media (e.g., 
de Gouveia Dal Pino, 
Birkinshaw, \& Benz 1996),
de Gouveia Dal Pino \& Birkinshaw 1996), 
momentum
exchange processes with the environment (Chernin, Masson, de Gouveia Dal Pino, \& Benz 1994), jet interactions 
with ambient clouds (de Gouveia Dal Pino 1999), and the effects of magnetic 
fields (e.g., Cerqueira, de Gouveia Dal 
Pino, \& Herant 1997, Cerqueira \& de Gouveia Dal Pino 1999). SPH is a 
Lagrangean, gridless approach to hydrodynamics,
in which the equations of continuity, momentum, and energy are solved
explicitly in time. Davies et al. (1993) have performed detailed comparisons
of SPH calculations with those based on more traditional, finite-difference
methods. In such tests, the two codes produce similar results 
although the spatial
resolution of 3D-SPH may be sometimes inferior (see, e.g., 
de Gouveia Dal Pino 1999 and references therein for a detailed 
discussion of the basic assumptions of the method 
and accuracy criteria).

In this preliminary investigation of the giant HH jets, we have neglected the effects of magnetic fields. Lately, these have been the object of intensive numerical study in HH jets by several groups (see, e.g., 
de Gouveia Dal Pino \& Cerqueira 1996, Stone, Xu \& Hardee 1997,
Cerqueira, de Gouveia Dal Pino \& Herant 1997, Frank et al. 1998, 
Gardiner et al. 2000, O'Sullivan \& Ray 2000, Stone \& Hardee 2000, Cerqueira \& de Gouveia Dal Pino 2000), despite the uncertainties related to their real strengths and topology. These studies have revealed that, while magnetic fields with reliable intensities and  geometries generally improve the collimation of the flow and may significantly affect their detailed structure, particularly behind the shocks,  they cause a minor influence on the dynamical structure of the beam  in comparison to pure hydrodymamical calculations. Thus as in the present analysis we are focusing on the deceleration behaviour  of the giant outflows, we may expect that the general results and conclusions of the this work will not be very much affected by the inclusion of magnetic fields.

The computational domain  is a three-dimensional, 
rectangular box, taken to represent the ambient medium. Within our Cartesian
coordinate system, the box has dimensions 
\hbox{$-50\,R_j\,\le\,x\,\le 50\,R_j$},
\hbox{$-9\,R_j\,\le\,y\,\le 9\,R_j$},
\hbox{$-9\,R_j\,\le\,z\,\le 9\,R_j$}, where $R_j$ is the initial jet radius
(and $R_j$ is the code distance unit).
The origin of the coordinate system lies at the center of the box. The
jet is continuously injected at \hbox{$(-50\,R_j,0,0)$} and subsequently flows
in the $x$-direction. Within the box, 330,000 particles are initially 
distributed uniformly in space. The particles are smoothed out in space and the fluid quantities are derived by employing a
spherically symmetric kernel function of width $h$. As in previous work, $h$ was chosen 
initially to be $0.25\,R_j$ and $0.50\,R_j$ for the jet and ambient particles,
respectively. The kernel width was then allowed to vary in space and time
during subsequent evolution, in order both to optimize resolution in
high-density regions, and to economize in the sparse, outer cocoon. (See also
de Gouveia Dal Pino \& Benz 1994; de Gouveia Dal Pino \& Birkinshaw 1996). The 
code advances all quantities in
time with a second-order, Runge-Kutta-Fehlberg integrator, and employs the
Newman-Ritchmyer artificial viscosity to handle shocks. Outflow boundary
conditions are assumed at the walls of the box.

We treat both the jet and the ambient gas as a fully ionized fluid that obeys
an ideal equation of state with \hbox{$\gamma = 5/3$}. Radiative cooling is
due to collisional excitation and recombination. We adopt a local prescription,
employing the cooling function of Katz (1989) for a gas of cosmic composition.
For temperatures below $10^4$~K, the assumption of complete ionization fails,
and the density becomes optically thick to its own cooling radiation. Rather
than deal with these complications, we have simply forced the cooling to zero
in this regime. A consequence of this adiabatic approximation is that we
underestimate the degree of clumping in relatively low-temperature regions. On
the other hand, we have also neglected the ambient magnetic field, which should
provide much of the supporting pressure in this regime (Hollenbach \& McKee
1979). Our treatment of postshock clumping, while insufficient for a detailed
study of the working surfaces (e.g., Blondin et al. 1990), should be adequate
for the present purpose of describing the larger-scale momentum transfer
between the jet and its environment.

The velocity of the jet at the circular inlet is given by 
\begin{equation}
v_o(t)\,=\,v_j\,+\,\Delta v\ {\rm sin} 
 \left({\frac{2\pi t}{P}}\right)\,\,. 
\end{equation}
Here $v_j$ is the mean jet speed, $\Delta v$ the amplitude of velocity 
variations, and $P$ their period. The models are also characterized by 
three additional,
nondimensional parameters: (a) the initial density contrast between the jet
and the ambient medium, \hbox{$\eta\,\equiv\,n_j/n_a$}, (b) the mean ambient
Mach number, \hbox{$M_a\,\equiv\,v_j/c_a$}, where $c_a$ is the ambient sound
speed, and (c) the pressure ratio at the inlet, 
\hbox{$\kappa\,\equiv\,p_j/p_a$}.

\section{Numerical Results}
\subsection{Pressurized Jets}
We have chosen the parameters of our study to be representative of observed
conditions within the giant Herbig-Haro flows. Where data were unavailable,
we utilize observations of the shorter, contiguous jets. As in previous studies, 
we fixed $n_j$,
the input jet number density, at 600~cm$^{-3}$, a figure representative of 
estimates derived through shock 
modeling of the emission lines
(e.g., Raymond et al. 1994). We further set the density contrast $\eta$
equal to 3 in all our runs. Previous simulations for steady jets indicate that
this figure yields a working surface for the leading bow shock that mimics 
the spatial clumping 
of such well-studied shocks as
HH~1 and 2 (Blondin et al. 1990; de Gouveia Dal Pino \& Benz 1993). Note that
a significantly lower value of $\eta$ (e.g., $\eta < 1$) results in a large,
inflated cocoon behind the leading shock, while a much higher figure
(\hbox{$\eta > 10$}) yields a dense, bullet-like working surface.  

As our first numerical experiment, we consider a jet that is in pressure
equilibrium with its surroundings, i.e., for which \hbox{$\kappa =1$}. 
As in previous works, we
choose the ambient temperature to be \hbox{$T_a\,=\,10^4\ {\rm K}$}, so that
\hbox{$c_a\,\equiv\,(\gamma k_B T_a/{\bar m})^{1/2}\,\simeq \,17\,{\rm km~s}^{-
1}$}.
Here, $k_B$ is Boltzmann's constant and $\bar m$ the mean mass per particle.
The corresponding temperature within the jet is
\hbox{$T_j\,=\,T_a\,\eta^{-1}\, \simeq \,3300\,{\rm K}$}, while the sound speed 
is
\hbox{$c_j\,=\,c_a\,\eta^{-1/2}\,=\,9.6\,{\rm km}^{-1}$}. The temperature $T_j$
is considerably lower than that in the postshock cooling regions that emit the 
observed optical emission lines, and should rather be viewed as an average 
between widely spaced shockfronts. In addition, our adopted $T_a$ is far greater
than any actual kinetic temperature within a molecular cloud. Nevertheless, our
imposition of pressure equilibrium is reasonable, since observed jets spread
much more slowly than their associated Mach angles (Ray \& Mundt 1993). How 
this collimation is actually achieved remains an open question. If, as is
widely believed, the collimating agent is the pinching force from a toroidal magnetic field component 
(e.g., Pudritz et al. 1991; Cerqueira \& de Gouveia Dal 
Pino 
1999, 2000; Gardiner et al. 2000), then our ambient pressure may
be considered a surrogate for this effect. We discuss in the next subsection
the consequences of relaxing the assumption of pressure balance.

We set both the jet's mean velocity $v_j$ and amplitude $\Delta v$ to
100~km~s$^{-1}$. These figures are appropriate for giant flows such as HH~34
(Devine et al. 1997). The mean ambient Mach number is therefore 
\hbox{$M_a\,=\, v_{j}/c_a \, = \, 6$},
and the 
maximum ambient and jet Mach numbers are $M_{a,max} =
v_{j,max}/c_a \simeq 12$, and
$M_{j,max} = v_{j,max}/c_j \simeq 21 $, respectively.
Also representative is the jet period, \hbox{$P\,=\,760$}~yr. The flow is
injected into a circular inlet of radius \hbox{$R_j\,=\,10^{16}$}~cm. The
latter value is about a factor of 10 greater than the widths of jets much
closer to the driving stars, as seen in HST images (e.g., 
Reipurth \& Raga 1999).
The more extended bow shocks in giant jets may indicate that the underlying
flows are also broader, but there has been no quantitative study of this 
point. In any case, the jet width at the inlet does affect the detailed pattern
of radiated emission, but not the gross dynamics of the jet-cloud interaction
(de Gouveia Dal Pino \& Benz 1993, Stone \& Norman 1993). Note finally that the 
fiducial time unit in our study
is that required for an ambient sound wave to traverse the jet radius. With
the indicated choice of parameters, this unit is \hbox{$t_d \,= R_j/c_a \, 
\simeq \,190$}~yr.
    
Figure 1 displays these results at a time 
\hbox{$t\,=\,22 t_d \, \simeq 
\, 4200$}~yr
after the initial injection 
(which corresponds  to the beginning of a new 
period with 
the injection of a new pulse at the inlet). 
At this point, the jet has propagated for
$100 \, R_j$, or almost 0.3 pc. The upper and lower panels show, respectively,
the density contours and velocity vectors in the midplane \hbox{($z = 0$)}. The
leading pulse or $working$ $surface$ is just leaving the right-hand border in 
both panels. This sharp
feature is followed by 4 others, including the one just forming near the inlet.
Since $P\,\approx\,4\,t_d$, 
there has been time for 7 pulses to 
form through steepening of the input sinusoidal velocity profile. 
The first 3 of 
these
have overtaken the leading one and merged with it. Each new pulse at the left
forms within a time of about $3\,t_d$. This figure agrees with that 
previously obtained by Raga et al. (1990) and Raga \& Canto (1998) in their 
analysis of steepening 
within a one-dimensional, sinusoidal injection profile.

The total propagation time for the HH~34 flow, from the star itself to the
last visible shocks is about $10^4$ ~yr (Devine et al. 1997).
Such a period may be brief compared to the expected 
evolutionary time scale of the central source, whether it is a true protostar
or a deeply embedded pre-main-sequence object. Thus, any results from 
simulations in which the leading pulse still plays a role are transients that
should be viewed in this light. 
It is important to establish, for example,
that the basic pattern seen in Figure 1 still holds even after the first
bow shock has left the computational domain. Our results indicate that this is
the case. 

To illustrate the point, Figure 2 shows results from two different epochs,
\hbox{$t\,=\,22\,t_d$} (as before) 
and 
\hbox{$t\,=\,24\,t_d\, \simeq 4600$}\,yr. 
The top two panels show the variation
of longitudinal velocity $v_x$ along the jet's central axis, while the bottom
two display the run of central density. We see that the velocity profile
exhibits the saw-tooth pattern familiar from earlier, one-dimensional studies
of free-streaming oscillating jets (e.g., Raga \& Kofman 1992). At the later 
time, each embedded peak shifts to the
right, but the overall pattern remains. The central density has pronounced 
spikes at each velocity minimum, i.e., just behind the leading shock within 
each pulse.
\footnote {The pulses formed by the steepening of velocity
in a variable jet consist of two shock fronts moving in tandem (e.g., Hartigan
\& Raymond 1993). Our simulations only resolve the two fronts within the
first few pulses, where they appear as breaks within the steep velocity
declines in Figure 2.}

The lower left panel of Fig. 2 ($t \,  = \, 22 t_d$) also shows that, 
in the beginning of each 
period, when a new pulse forms at the jet
inlet, its density is initially high, 
due to the efficient radiative cooling of the
shocked material,  
and then decreases as the pulse propagates. (At  $t = 22 t_d$, 
the new pulse has an initial density $ \sim 65 n_a$ 
which  decreases to $\sim 9 n_a$  at $t = 24 t_d$, in the lower right 
panel.)
In fact, at both epochs, the downward sharp density 
spikes fall off
in the downstream direction. This effect has been studied 
in previous numerical  work 
(see, e.g., de Gouveia Dal Pino \& Benz 1994, 
V\"olker et al. 1999),
and is consistent with measurements of electron density decay 
with distance in HH jets (e.g., Reipurth 1999). The initial 
velocity
differences impressed on the flow at the inlet are gradually erased as faster
material catches up with slower, upstream gas thus producing the saw-tooth 
pattern. 
The peak velocities are falling off with
distance in a nearly linear fashion (by
a factor $\sim$ 2 within $\sim $ 0.3 pc).

In our simulation, we also
find that the propagation speed of 
each pulse (or internal working surface, WS), 
$v_{\rm WS}$, diminishes,
although at a slower rate. Figure 3 plots this speed at the same two epochs.
Here we derived $v_{\rm WS}$ by differencing the position of the density
spike at closely spaced times. The segmented nature of the curves 
results
from the fact that this velocity has been computed only for some 
specific positions of 
the density
spikes. Note that the severe, final drop in $v_{\rm WS}$ at 
\hbox{$t\,=\,22\,t_d$} represents the deceleration of the leading bow shock,
as it sweeps aside warm, ambient gas. In fact, 
its propagation velocity which is
initially
$v_{WS,o} \simeq v_{j,max}/[1 + \eta^{-1/2}] \simeq \, 7.5 \, 
c_a \sim $ 126 km s$^{-1}$ (e.g., de Gouveia Dal Pino \& Benz 1993), 
decreases to $v_{WS,o} \sim 3.5 \, c_a$  at $t = 22 t_d$, just  before it 
leaves the computational domain, as indicated 
in Fig. 3 (top panel). The results above are
in qualitative 
agreement with the observed kinematics of HH34 system for which a 
deceleration factor $\sim 1.5$ has been  observed within the 
same distance.

Fig. 3 also depicts the calculated propagation velocity of the internal 
pulses along the jet's central axis which was estimated under the 
assumption that each pulse traps only a small 
amount of 
gas 
between its two shocks (and eject most of the material sideways). 
In this case, the motion of an internal pulse 
is obtained from the balance between the pre and post 
pulse  momentum fluxes (neglecting the gas pressures up- and 
down-stream) 
which gives (see, e.g., Raga \& Kofman 1992, de Gouveia Dal Pino \&
Benz 1994):
\begin{equation}
v_{WS,cal} \sim 
\frac{ (\rho_l/\rho_r)^{1/2} v_l + v_r }{ (\rho_l/\rho_r)^{1/2} +1} 
\end{equation}
\noindent
where $v_l$ and $v_r$ are the flow velocities, and
$\rho_l$ and $\rho_r$ are the densities
 immediately up- and down-
stream of each  pulse, respectively. Substituting in the equation
above  the measured values from the 
simulation for $\rho_l/\rho_r$, $v_l$, and $v_r$ for each pulse (see Fig. 2) 
we have 
derived the dashed  curves of Fig. 3. 
Eq. (2) which is valid for 
one-dimensional shocks under the conditions described above 
should be applicable 
to the jet axis of Fig. 1, where the shock velocity is mostly parallel to 
the jet axis.  The slight disagreement of the dashed 
curves in Fig. 3 with the propagation velocities 
measured in the
simulations (solid curves) is 
caused by the 
fact that some amount of momentum is escaping sideways, even from the
central region.

Figure 4 displays the variation of 
momentum flow at
the time \hbox{$t\,=\,24\,t_d$}. 
The dashed curves are
profiles of the momentum flux, $\rho\,v_x^2$, shown as a function of the 
transverse coordinate $y$. 
Each profile is computed at the $x$-position 
where the flux asymptotes to zero. 
Comparing with Figure 2, we see how the
momentum flux rises at the position of each pulse, and then falls in between.
The actual peak value of $\rho\,v_x^2$ systematically declines, an indication
that the jet is losing momentum.

The solid curves in the panel show the momentum transport rate, represented by
the quantity \hbox{$\int_0^y\!\rho\,v_x^2\,y\,dy$}. Since the flow is nearly
symmetric about the central axis, we may replace $y$ by 
\hbox{$r/\sqrt{2}\,\equiv\,[(y^2 + z^2)]^{1/2}$} in the integral, which is then
proportional to the momentum per unit time crossing within a circle 
of radius $r$
at each $x$-position. The curves are shallowest between pulses, where the least
momentum flow occurs. Conversely, they are deepest at the pulses themselves. 
Their
depth at the pulse positions declines from the
first pulse (near the inlet) to the last one shown. 
In other words, {\it the full momentum transport rate is lower
at the far end than at the inlet}. 
The excess momentum must be taken up by
whatever mass leaves the box, i.e., by the external medium.

\subsection{Overpressured Jets}  
We next consider a jet that is injected with significantly higher pressure than
its surroundings. Retaining \hbox{$\eta\,=\,3$} and 
\hbox{$n_j\,=\,600$}~cm$^{-3}$, we now lower the ambient temperature $T_a$ to
$10^3$~K. We further take the initial temperature of the jet itself to be
\hbox{$T_j\,=\,10^4$}~K. Thus, the new pressure ratio is 
\hbox{$\kappa\,=\,\eta\,T_j/T_a\,=\,30$}. 
Our input velocity profile is still
given by equation (1), with \hbox{$v_j\,=\,\Delta v\,=\,100$}~km~s$^{-1}$, but
we lower $P$ slightly to 600~yr. Since the ambient sound speed is now only
5.3~km~s$^{-1}$, the corresponding mean ambient Mach number is higher: 
\hbox{$M_a\,\equiv\,v_j/c_a\,=\,19$}, 
$M_{a,max} =
v_{j,max}/c_a \sim 38$,  and 
$M_{j,max} = v_{j,max}/c_j \simeq 12 $, with  $c_j \simeq 16.6 $ 
km s$^{-1}$.
 The fiducial crossing time is now raised
to \hbox{$t_d = R_j/c_a \, \simeq \,600$}~yr. Thus, we now have 
\hbox{$P\,=\,t_d$}.

Figure 5 shows density contours in the midplane for
\hbox{$t\,=\,2\,t_d\, \simeq\,1200$}~yr,
\hbox{$t\,=\,4\,t_d\, \simeq\,2400$}~yr, and 
\hbox{$t\,=\,6\,t_d\, \simeq\,3600$}~yr.  
The later frame shows the jet just before the leading bow shock exits
the computational boundary.
In comparison with the upper panel of
Figure 1, we see that the flow has expanded more rapidly in the transverse
direction. One point in common is that, although there has been time to form
7 pulses, only 4 are visible.
As before, 3 of the pulses originally formed have overtaken the
leading bow shock and merged with it.
\footnote{
We note that  in this particular case of  the overpressured jet where the rapid transverse expansion
could introduce undesirable boundary effects in the evolution of the flow, we made  
previous tests by simulating a shorter jet propagating in
a computational domain with much larger transverse boundaries and 
an agreement better than 15\% was found
with the results of  the narrower boundary simulation.}

The variations in velocity and density along the central axis are displayed,
respectively, in the top and bottom panels of Figure 6. Here we chose as
representative times \hbox{$t\,=\,6\,t_d\, \simeq \, 3600$}~yr, 
and \hbox{$t\,=\,9\,t_d\, \simeq \,5400$}~yr. Although
the overall velocity profile still approximates a saw-tooth pattern, the peak
value of $v_x$ declines more slowly than before, except for the steep drop 
associated with the leading  bow shock at the early time. The density spikes
show no falloff at this epoch. Note especially the large pileup in mass just
behind the leading bow shock. By \hbox{$t\,=\,9\,t_d$}, the density spikes
again decline, although somewhat erratically. The propagation speed of the
pulses also displays no simple pattern (Fig. 7). 
The leading  bow shock is slowing down by nearly 
the same rate as the jet flow, from  an initial value 
$v_{WS,o}  \simeq \, 22 c_a \sim $ 126 km s$^{-1}$ to 
$v_{WS,o} \simeq  14 c_a \simeq$ 80 km s$^{-1}$
 at $t = 6 \, t_d$, but the  
velocity of propagation of the internal pulses, $v_{WS}$, approximates 
the peak velocities of Fig. 6.
We note also in Fig. 7 that the predicted propagation velocities of the 
internal pulses in the jet axis by eq. 2 (dashed lines) are now in 
less
agreement with the 
measured values from the simulations than in the pressure-confined jet,
thus suggesting that the condition of a 
negligible amount of gas being trapped between  the shocks of each
pulse is less 
sustained here. 

Figure 8 displays the variation of 
momentum flow at
the time \hbox{$t\,=\,9 \,t_d$}. 
As in Fig. 4, the dashed curves are
profiles of the momentum flux, $\rho\,v_x^2$, 
shown as a function of the 
transverse coordinate $y$, and the 
solid curves show the 
momentum transport rate
at each $x$-position. 
Again, the depth of the profiles rises 
and falls at pulse and interpulse
locations, respectively,
and the curves very near the pulses become shallower
with increasing $x$ (although more irregularly than 
in the case of the pressure-confined jet) thus indicating that  
momentum transport rate is lower
at the far end than at the inlet.

The smaller deceleration rate of the overpressured jet
is primarily related  
to the lesser  ability of the cold ambient medium in slowing down
this higher Mach-number jet. 
Besides, although, like the
pressure-confined jet of Fig. 1, it is progressively
loosing momentum sideways behind each internal pulse which
causes  deceleration, 
the jet is also undergoing an expansion specially near the inlet 
(due to the 
overpressure). This causes the development of
crossing shocks that propagate from the edge of the
beam inwards and produce partial refocusing of the sideways expanding
material to the axis thus inhibiting the deceleration process.


\subsection{Steady Jet}
In order to test the importance 
of the primary impact of the 
pressure confined jet into the ambient medium, we also performed a 
numerical experiment considering a steady jet 
with the same input conditions of the
pressure-confined jet of Fig. 1,  but injected with a constant velocity
$M_a = v_{j}/c_a \simeq 12$. 
Figure 9 depicts this steady jet just before it exits the computational domain, 
at $t \simeq 11 t_d$, and Figure 10 displays the propagation velocity of the
leading working surface at the jet's central axis as a function of the distance. 
We note that, 
instead of being
decelerated by the impact with the 
warm ambient medium, the steady jet is first  
accelerated
and then reaches a constant velocity regime. The dashed curve gives the 
predicted propagation velocity of the leading working surface in the 
case that the jet radius at the head ($R_h$) is $R_h \simeq  R_j$, i.e., 
$v_{WS,o} \simeq 
v_{j}/[1 + (\eta \alpha ^{-1/2}] \simeq \, 7.5 \, 
c_a \sim $ 126 km s$^{-1}$, 
where $\alpha = (R_j/R_h)^2 \simeq 1$ (e.g., de Gouveia Dal Pino \& Benz 1993).
The acceleration effect in Fig. 10 has been previously reported 
for radiatively cooling, overdense 
steady-state jets (de 
Gouveia Dal Pino  \& Benz 1993) and, according to the
equation  above,  is directly related to the progressive decrease of the 
radius  at the head as the jet propagates downstream (see Fig. 9).
This result indicates that the deceleration observed in 
giant jets $cannot$ be simply  attributed to  the impact of the 
leading working surface with the ambient 
medium. It is due 
instead to the momentum transfer by the gas expelled sideways by 
the multiple  
working surfaces to
the ambient medium as we have seen in the previous paragraphs.

\section{Discussion and Conclusions}
We have performed  fully three-dimensional simulations of
 overdense, radiatively cooling jets
evaluated for a set of parameters which are appropriate
to giant protostellar jets. 

Modeling  jets with long-period (P $\sim$ several hundred years)
and large amplitude sinusoidal velocity variability at
injection ($\Delta v =$ mean jet flow velocity), and allowing them to
travel over a distance well beyond the source,
the injected sinusoidal flow velocity
profile steepens into a saw-tooth pattern, in agreement with previous
theoretical work (e.g., Raga \& Koffman 1992, V\"olker et al. 1999),
developing a chain of travelling internal pulses. 
For a 
pressure-confined jet (Figs. 1 to 4), 
the density spikes 
behind each pulse decrease as they propagate away from the source.
Also, we find that the
peak velocities of the  saw-tooth flow and the propagation 
velocity of the internal pulses fall off with 
distance by 
a factor $ \lesssim$ 2 within $\sim 0.3$ pc, which is 
consistent with the observed kinematics of the HH34 system 
within the same distance.
This deceleration is slower for an overpressured jet 
with similar input conditions but higher Mach number
(Figs. 5 to 8). 
The analysis of the momentum flux along the flow and the total momentum rate 
crossing laterally the jet radius indicates that the deceleration in
the pulsed giant jets  
is mainly caused by systematic momentum transfer sideways
by the traveling pulses
to the ambient medium.

We note that, in a previous work (Chernin et al. 1994), we have investigated
 numerically the possible processes of  momentum transfer from  astrophysical  jets to the ambient medium and
found that, while 
low Mach number (smaller than or equal to 6) jets slow down rapidly because
they entrain turbulently ambient  material along their sides, 
high Mach number beams, like the HH jets studied here, on the other hand,
 transfer  momentum to the ambient medium 
principally at the working surfaces (i.e., through "prompt" entrainment).  The results of the present work  are, therefore, consistent
with this former analysis.

Also, the simulation of a pressure-confined steady-state jet (Figs. 9 and 10) 
with the same 
initial 
conditions of the pressure-confined pulsed jet above 
has shown that the steady jet experiences some acceleration, 
which is an additional  indication that 
the primary source of the deceleration observed in the giant jets could not be 
attributed
to braking of the leading bow shock against the ambient medium.

In a recent
work, Cabrit \& Raga (1999) utilizing a very 
simple analytical model have hypothesized  
that the observed jet deceleration in 
the HH 34 system could be possibly the result of drag forces of the ambient 
medium on individual jet knots arising from fragmentation 
of a time-variable, precessing jet.
As we argued earlier, the symmetric pattern of deceleration in both lobes
of the jet indicates that the process is rather internal to the flow. 
Our three-dimensional simulations show, in fact, that deceleration
can be
naturally produced in jets formed by long-period,
large-amplitude velocity variations,
where momentum is 
progressively lost from the central axis and transferred 
sideways by the travelling pulses.

Finally, we note that, for the sake of simplicity and also 
because observations offer only average estimates, 
we have assumed  in this study an initially homogeneous ambient medium, although
we could expect some stratification in density and temperature,
and even in pressure, 
of  the environment surrounding the giant flows.
Previous  numerical studies involving stratified ambient media  have 
been done both in
the context of 
extragalactic jets 
(see., e.g.,   Wiita et al. 1990, Hardee et al. 1992), and the shorter protostellar jets (see, 
 e.g., de Gouveia Dal Pino, Birkinshaw, \& Benz 1996, de Gouveia Dal Pino \& Birkinsahw 1996),
but 
in the future, a more complete analysis of the giant flows must include 
a stratified ambient medium for a complete investigation of its effects upon the structure and long term evolution of the flow over the cloud.
For example, for an overdense jet, as long as the pressure is maintained constant, one could expect that  ambient  density and temperature variations by,  say
a factor of two or so, while having a minor effect on the jet dynamics,
they could more
significantly affect the radiative emission structure of the shocks  
(due to the strong dependence of
the radiative cooling distance  with the shock velocity and thus with the density ratio between the jet and the ambient medium; see, e.g., de 
Gouveia Dal Pino \& Benz 1993). 

\acknowledgements
This work has benefited from several valuable and enthusiastic 
discussions  and profitable comments from S. Stahler, F. Shu,
and A. Raga, 
during my visit to to the University of California at Berkeley.
I am particularly indebeted to S. Stahler who suggested
the original idea of this work.
This work has been partially supported by grants of the
Brazilian agencies FAPESP and CNPq.

\newpage

\centerline {\bf FIGURE CAPTIONS}

\noindent {\bf Figure 1:}  Mid-plane density contour (top)
and  velocity field distribution (bottom)
for a pressure-confined
 radiatively cooling, pulsed jet propagating into a warm ambient 
medium with $T_a \sim 10^4 $ K. The sinusoidal velocity variability at 
injection has a period 
$P \sim 4 t_{d} \sim 764 $ yr
(where $t_d = R_j/c_a 
\approx 190$ yr, and $R_j = 10^{16}$ 
cm), a mean speed 
$v_j = 100 {\rm km s^{-1}}$, and  an amplitude
$\Delta v = 100 {\rm km s^{-1}}$. 
The time depicted is 
$t/t_d \, \sim \, 22 \, \sim 4200$ yr, 
and the jet
 has propagated for $100 \, R_{j} \, \sim \,0.3$ 
pc. 
The initial
conditions for the jet are: $\eta=n_j/n_a=3$, $n_a= 200$ cm$^{-3}$, 
$M_{a,max} = 12$, and 
pressure equilibrium in the 
jet at the inlet, $\kappa=p_j/p_a = 1$.
The distances are in units of $R_j = 10^{16}$ cm.
The density lines are separated by a factor of 1.3 and
the density scale covers the range
$\sim$ (1.3$\times 10^{-2}$   -  $0.7\times 10^2) n_a$.

\noindent {\bf Figure 2:} Flow velocity (top) and density (bottom) 
profiles along the jet axis for the jet of Fig. 1,
at $t/t_d = $ 22 (left), and $t/t_d = $ 24 (right). 
The flow velocity is in units of
the ambient sound speed,  $c_a  \simeq $ 16.6 km s$^{-1}$, the density 
is in units of the ambient density $n_a = $ 200 cm$^{-1}$, and the 
distance 
is
in units of $R_j = 10^{16} $ cm.

\noindent{{\bf Figure 3:} Propagation velocity of the leading and internal 
working surfaces, 
$v_{WS}$ (in units of $c_a \, = \, 16.6 km s^{-1}$), for the jet of Fig. 
1 at $t/t_d \simeq $ 22, and 25. The distance 
is
in units of $R_j = 10^{16} $ cm. 

\noindent {\bf Figure 4:} 
Longitudinal component of the momentum flux density 
(dashed lines) and the integrated momentum rate distribution 
(solid lines)
for the jet of 
Figs. 1 to 3, 
at $t/t_d \simeq $ 24, in
different positions along the flow. At the positions:
$x/R_j =$ -29, -7, 16, and 36, the lines trace the distributions 
profiles immediately behind the pulses; at the other positions, they 
trace the distributions at regions between the pulses.
The momentum flux density, and the mometum rate
scales can be calibrated 
using the markers in the top region of each plot - the marker for the 
momentum flux density 
corresponds to 
$\sim 2.9 \times 10^{-18}$  g cm$^{-3}$  (km s$^{-1})^2$, and
the marker for momentum rate corresponds to
$\sim 1.4 \times 10^{-6} M_{\odot} $ yr$^{-1}$ km s$^{-1}$.

\noindent{{\bf Figure 5:} Mid-plane density contour evolution
of an overpressured
 radiatively cooling, pulsed jet propagating into an ambient 
medium with $T_a \sim 10^3 $ K.  The sinusoidal velocity variability at 
injection has a period 
$P \sim 1 t_{d} \sim 604 $ yr
(where $t_d = R_j/c_a 
\approx 604$ yr, and $R_j =10^{16}$ 
cm), a mean speed 
$v_j = 100$ km s$^{-1}$, and  an amplitude
$\Delta v = 100$ km s$^{-1}$. The times depicted are 
$t/t_{d} \, \sim \, 2 \, \sim \, 1200$ 
yr, 
$t/t_{d} \, \sim \, 4 \, \sim \, 2400$ yr,
 and $t/t_{d} \, \sim \, 6 \, \sim \, 3600$ yr.
The initial
conditions for the jet are: $\eta=n_j/n_a=3$, $n_a= 200$ cm$^{-3}$, 
$M_{a,max} \sim  38$, and a jet to ambient 
pressure ratio
at the inlet $\kappa=p_j/p_a \sim 30$, with  an initial jet temperature
$T_j = 10^4$ K.
The distance is in units of $R_j = 10^{16}$ cm.
The density lines are separated by a factor of 1.3 and
the density scale covers the range
$\sim$ (1.1$\times 10^{-2}$   -  $0.51\times 10^2$) $n_a$ at 
$t/t_{d} \, \sim \, 2$,
$\sim$ (1.2$\times 10^{-2}$   -  $0.5\times 10^2$) $n_a$ at
$t/t_{d} \, \sim \, 4$,
and $\sim$ (1.1$\times 10^{-2}$   -  $0.99\times 10^1$) $n_a$
at $t/t_{d} \, \sim \, 6$.

\noindent{{\bf Figure 6:} Flow velocity (top) and density (bottom) 
profiles along the jet axis for the jet of Fig. 5,
at $t/t_d \sim $ 6 (left), and $t/t_d \sim $ 9 (right). 
The flow velocity is in units of
the ambient sound speed,  $c_a  \simeq $ 5.25 km s$^{-1}$, the density 
is in units of the ambient density $n_a = $ 200 cm$^{-1}$, and the 
distance 
is
in units of $R_j =$ 10$^{16} $ cm.

\noindent{{\bf Figure 7:} 
Propagation velocity of the leading and internal working surfaces, $v_{WS}$ (in 
units of $c_a  = 
$ 5.25  km s$^{-1}$), for the jet of Fig. 5 at $t/t_d \simeq $ 6, and 9. The 
distance 
is
in units of $R_j =$ 10$^{16} $ cm.

\noindent{{\bf Figure 8:} Longitudinal component of the 
momentum flux density 
(dashed lines) and the integrated momentum rate distribution 
(solid lines)
for the jet of 
Figs. 5 to 7, 
at $t/t_d \simeq $ 9, in
different positions along the flow. At the positions:
$x/R_j =$ -25, 2.5, and 35, the lines trace the distributions 
profiles immediately behind the pulses; at the other positions, they 
trace the distributions at regions between the pulses.
The momentum flux density, and the momentum rate
scales can be calibrated 
using the markers in the top region of each plot - the marker for the 
momentum flux density 
corresponds to 
$\sim 1.7 \times 10^{-18}$  g cm$^{-3}$  (km s$^{-1})^2$,
and the marker for the momentum rate corresponds to 
$\sim 1.2 \times 10^{-9} M_{\odot} $ yr$^{-1}$ km s$^{-1}$.

\noindent {\bf Figure 9:}  Mid-plane density contour (top) and
velocity field distribution (bottom) for a pressure-confined
 radiatively cooling, steady-state jet propagating into an ambient 
medium with $T_a \sim 10^4 $ K with constant 
velocity at injection $v_j \simeq 200$ km s$^{-1}$, and $M_{a} = 12$. 
The time depicted is $t/ t_d \simeq$ 11
$\simeq $ 2101 yr, (where $t_d = R_j/c_a 
\approx 190$ yr, and $R_j = 10^{16}$ 
cm), and the jet
 has propagated for 
$\sim 100 R_j\, \sim  0.3$ pc. 
The initial
conditions for the jet are: $\eta=n_j/n_a=3$, $n_a= 200$ cm$^{-3}$, 
and 
pressure equilibrium in the 
jet at the inlet, $\kappa=p_j/p_a = 1$.
The distances are in units of $R_j = 10^{16}$ cm.
The density lines are separated by a factor of 1.3 and
the density scale covers the range
$\sim$ (1.2$\times 10^{-1}$   -  0.4 $\times 10^{2}$) $n_a$.

\noindent{{\bf Figure 10:} Propagation velocity of the leading 
working surface 
$v_{WS,o}$ (in units of $c_a \, = \, 16.6$ km s$^{-1}$), for the jet of Fig. 
9. The distance 
is
in units of $R_j =$ 10$^{16} $ cm.

\end{document}